\documentclass{svproc}
\usepackage{url}
\usepackage{caption}
\usepackage{subcaption}
\usepackage{amssymb}
\usepackage{graphicx}
\usepackage[table,xcdraw,svgnames]{xcolor}

\begin{document}
\mainmatter              
\title{Data compression to choose a proper dynamic network representation}
\titlerunning{Data compression to choose a proper dynamic network representation}  
%
\author{Remy Cazabet\inst{1}}
\authorrunning{Remy Cazabet} 
%
%
\institute{Univ de Lyon, CNRS, Université Lyon 1, \\ LIRIS, UMR5205 Villeurbanne France\\
\email{remy.cazabet@gmail.com}}

\maketitle              

\begin{abstract}
Dynamic network data are now available in a wide range of contexts and domains. Several representation formalisms exist to represent dynamic networks, but there is no well known method to choose one representation over another for a given dataset. In this article, we propose a method based on data compression to choose between three of the most important representations: snapshots, link streams and interval graphs. We apply the method on synthetic and real datasets to show the relevance of the method and its possible applications, such as choosing an appropriate representation when confronted to a new dataset, and storing dynamic networks in an efficient manner.
\keywords{temporal networks, dynamic networks, link streams, information theory}
\end{abstract}
\section{Introduction}
The analysis of dynamic networks is an important topic of research in the network science community. With the ubiquity of digital data collection, more and more relational data becomes available with a temporal component. However, the way to handle and model such data is still a research question. As discussed in several articles in the state of the art \cite{holme2012temporal,latapy2018stream,cazabet2019challenges,gauvin2018randomized}, there are multiple ways to model the same original observations. 

In this article, we propose a method to choose an appropriate dynamic graph model, among the three following ones: Sequence of Snapshots, Interval Graphs and Link Streams. The method we propose is based on the principle of maximizing data compression, i.e., minimizing the network description's encoding length.

In section \ref{motivation}, we explain the rationale of our approach, and possible applications. In section \ref{encoding}, we introduce the computation of the encoding cost of a dynamic network, for each representation. Section \ref{experiemnts} present experiments on synthetic and real datasets. Finally, we conclude in section \ref{conclusion}.

\section{Context and Motivation}
\label{motivation}
Dynamic networks can be used to represent a variety of real-world phenomena, with widely different properties. For instance, some networks represent \textit{interactions} (e.g., e-mails, instant messages on Social Media, physical proximity, co-authoring in scientific networks, etc.), while some others represent persistent \textit{relations} (e.g., follower/followee relations on tweeter, friendship relations, active collaborations in scientific networks, etc.). It is clear that those two types of networks are of very different nature, and should be modeled and analyzed in different ways. But the difference between those two types is not as obvious as it might seem at first sight: some interactions have a duration (e.g., phone calls, physical proximity, etc.), while the nature of some collected data might be ambiguous (sentimental relations between teens, collaboration on a scientific project, etc.). 

Another source of difficulty is that data are often collected at a given temporal granularity, either for convenience or for technical constraints. For instance, scientific publications are often characterized by their publication year only, interaction logs might be rounded-up to the hour or even the day for privacy reasons, and large datasets such as friendship in Facebook are often collected at a low frequency, e.g., once a month or year.

For all these reasons, the choice of a model is often not as simple as knowing the nature of the studied data, but requires to look at the data properties.

\subsection{The different models of dynamic networks}
In this article, we will consider three types of dynamic network models, often used in the literature.
\begin{itemize}
    \item Snapshot (SN): The network is represented as a sequence of graphs. Each graph corresponds to a point in time or is made of the sum of all interactions over a period.
    \item Link Stream (LS): The network is a collection of edges, each identified by a pair of node and a point in time
    \item Interval Graphs (IG): The network is a collection of edges, and each edge exists over a given time interval, identified by its start and end times.

\end{itemize}

While these representations might appear unrelated at first sight, they are in fact able to represent the same original data as long as time is provided as a  \textit{discrete} value, which is the case in most practical situations. For instance, if time is represented as a POSIX time (timestamp), it can be considered discrete, since there is a countable number of possible values between any two POSIX time.

The best way to understand how the same data can be represented by these different models is to take a practical example: the SocioPatterns projet \cite{barrat2013empirical} has collected several physical interaction datasets between individuals in various contexts, such as schools or hospitals. Wearable sensors collect every face-to-face interaction at a high frequency between a group of subjects over an extended period of time, from a few hours to a few days. For practical reasons, data collection is made for the whole setting every 20 seconds, in a synchronous fashion. The publicly provided data therefore consists in a file containing all those captured interactions, as triplets $<T,U1,U2>$, with $T$ the timestamp of the interaction, $U1$ and $U2$ the face-to-face individuals. There are therefore several ways to interpret such data:
\begin{itemize}
    \item SN: Each 20s, a snapshot of all on-going conversations is captured. Each of these snapshots could be studied as a conventional static graph.
    \item LS: Each triplet is a link of the link stream, it corresponds to an observed interaction between individuals.
    \item IG: Individuals do not interact punctually but over periods of time, usually longer than 20s. Technical reasons force to capture the state of these persistent interactions periodically every 20s. To model more accurately the observed phenomenon, one should create continuous intervals for each pair of node interacting repeatedly every 20s over a period, lasting from the first to the last observation of the series.
\end{itemize}
Any of these interpretations is valid a priori, so the choice of using one instead of another is usually based on practical reasons, e.g., to apply a method that requires to have the data in one format or another. For instance, dynamic community detection algorithms assume a specific network format: Snapshots in most cases (e.g.,\cite{mucha2010community}), but also sometimes Link streams (e.g.,\cite{viard2016computing,matias2018semiparametric}) or Interval Graphs (e.g., \cite{cazabet2010detection,coscia2012demon}).

\subsection{Using encoding cost as a selection criterion}
The principle that the best description of data is the description that minimizes the cost of its representation can be found in several areas of science, from Occam's Razor to the Minimum Description Length \cite{grunwald2007minimum}(MDL) principle. 

For static networks, this principle could for instance be used to choose between a matrix representation, an edge-list representation and an adjacency list representation. For an unweighted, undirected network composed of $n$ nodes and $m$ edges, the cost (in bits) of a matrix representation is $n^2$ (a matrix of boolean values), while its corresponding representation as an edge list is $2m  \log_2(n)$ --encoding each edge requires to encode each of its 2 nodes. It means that if the graph is sparse, $m \ll n^2$, the edge list representation is the most efficient, and vice versa. The adjacency list representation is beyond the scope of this article, but relatively similar to the adjacency list. A first implication is that we can choose the most appropriate way to store the graph in memory given its properties $n$ and $m$, but, beyond this, it also provides hints on what can be done or not on this graph. For instance, in the community detection problem, methods using matrix factorization are little penalized by the density of the matrix, while an algorithm such as Louvain, designed for sparse graphs, only requires the neighborhoods of nodes, available in an adjacency list representation. Therefore, the best way to encode a graph also gives us hints on how to handle it.

Note that in this paper, we limit ourselves to the comparison of encoding scheme that depends only on the number of nodes, edges and temporal information, and not on other properties such as the degree distribution, that could also be optimized with techniques such as Huffman Coding. When dealing with dynamic graphs, we will also make the assumption in our representation that cumulated graphs of networks to represent are sparse, since this is the most common setting. We therefore propose representations which are extensions of edge lists rather than adjancency matrices.

\subsection{Applications}
The method introduced in this article is implemented in \textit{tnetwork}\footnote{https://tnetwork.readthedocs.io}, a python library to manipulate temporal networks. The first application is to automatically choose the most efficient in-memory representation for a temporal network loaded from a file containing triplets $<T,U1,U2>$, as is the case for temporal networks shared by the SocioPatterns project and those available on the \textit{Network Repository}\footnote{\url{http://networkrepository.com}} website\cite{networkrepo}.

The method is also used in the library to choose the most parsimonious representation when saving temporal networks created using the library, such as random temporal networks with community structure

Beyond these memory-related applications, knowing the most appropriate representation also tells the practitioner how to efficiently manipulate their data. For instance, if the snapshot representation is inefficient to represent a dynamic network, it is unwise to analyze each period as an independent snapshot, for instance computing centralities or detecting communities for each of them. Reciprocally, if the network is poorly represented as a link stream, it is unwise to apply methods expecting such a graph, e.g. the community detection method introduced in \cite{matias2018semiparametric}.

\section{Temporal network encoding cost}
\label{encoding}

\begin{figure}[h!]
  \centering
  
    \includegraphics[width=0.8\linewidth]{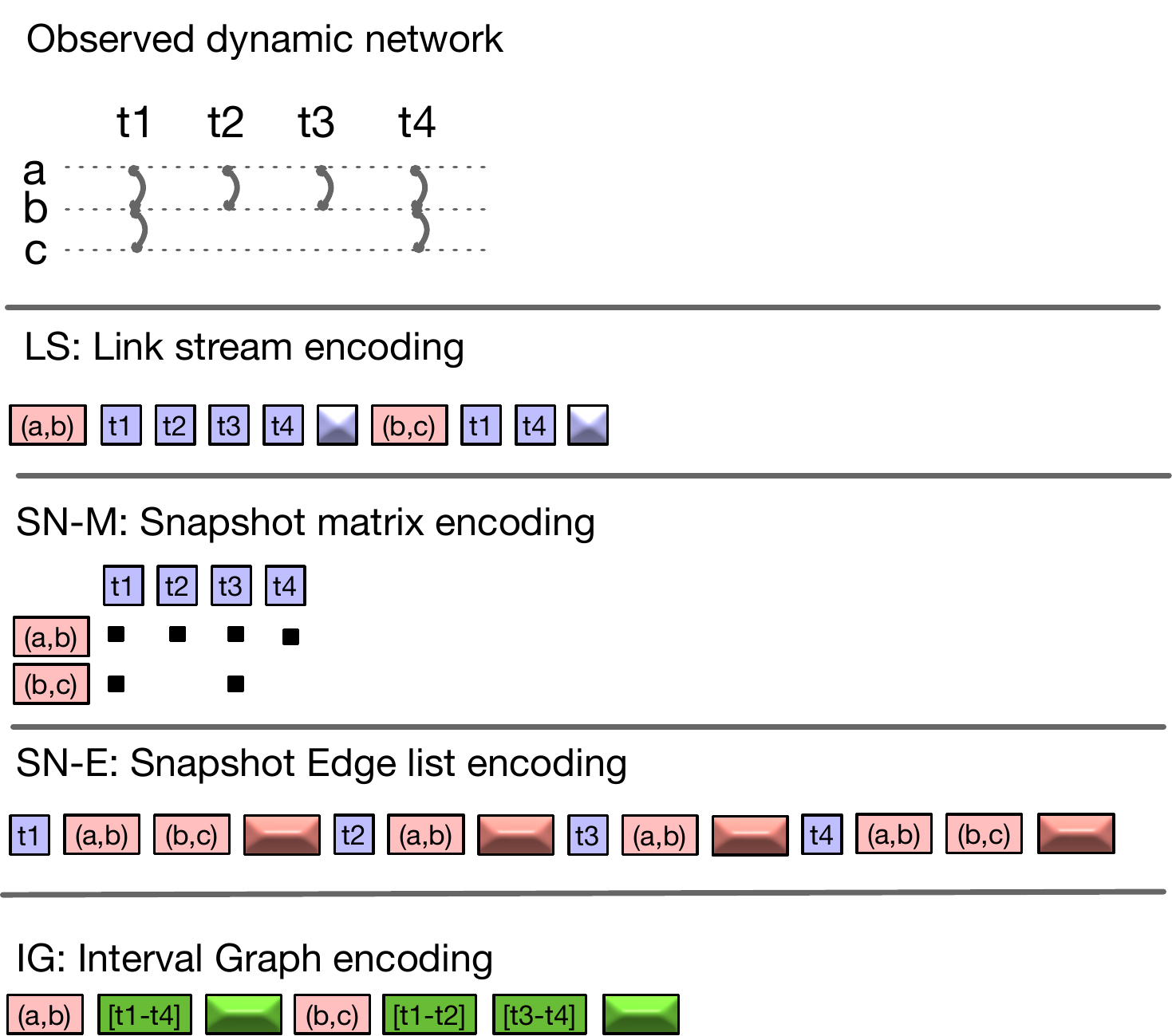}
  
  \caption{Illustration of the chosen encoding strategies. From an observed dynamic network, each strategy encodes node pairs (red) and temporal information (blue/green). Textured blocs correspond to STOP symbols to mark the end of a series of unknown length.  }
  \label{schema}
\end{figure}

Following \cite{latapy2018stream}, let's define our dynamic network as a link stream $L=(T,V,E)$, where $T$ is the list of possible times, $V$ the list of possible vertices, and $E$ triplets composed of two vertices and a time. We also define the aggregated graph $G^a(L)=(V,E^a)$, such as $E^a$ is the list of pairs of nodes (edges) that appear at least once in the link stream $L$.
The encoding cost of a given temporal network with a given representation depends mainly on three properties:
\begin{itemize}
    \item $m=|E^a|$ is the number of different edges to appear at least once
    \item $e=|E|$ is the total number of events, i.e., links in the link stream
    \item $t=|T|$ is the total number of different times during which at least one event occurs in the dynamic network.
\end{itemize}
We also need two partial encoding costs:
\begin{itemize}
    \item $I^t=\log_2(t)$ is the cost to encode one time information
    \item $I^m=2\log_2(|V|)$ is the cost to encode one node pair
\end{itemize}

The cost of encoding nodes themselves is a constant for a given network and is thus ignored in the rest of this paper.

\vspace{0.3cm}
\textbf{Link stream encoding}
For this encoding, we list, for each pair of nodes, the list of timestamps it appears in. The total encoding cost of a graph is:
\[
I^{ls}=mI^m+eI^t+mI^t
\]
Where the first element is the cost of encoding the edges, the second element encodes the timestamps, and the last encode stop sections to signal the end of a list of times.

We formalize below the cost of encoding a dynamic network using four representations. A summary of these representations can be found in Fig. \ref{schema}.
\vspace{0.3cm}

\textbf{SN encoding}
We consider two ways of encoding snapshot sequences, the first one to represent snapshots that have most of their edges in common, and the other one for snapshots that are few of their edges in common.

In this first snapshot representation, we encode data as a matrix, whose lines correspond to pairs of nodes and columns to timestamps. A unique bit is required to indicate if an edge appear at a given time or not.
\[
I^{SN_M}=mI^m+tI^t+te
\]
Where the first element is the cost of encoding the edges, the second element encodes the timestamps, and the last encode the bits of the matrix

In the second snapshot representation, each snapshot is represented as a list of pair of nodes, and timestamps are added at the start of every snapshot. This representation is equivalent to representing each snapshot as an edge list.
\[
I^{SN_E}=eI^m+tI^t+tI^m
\]
Where the first element is the cost of encoding the edges, the second element encodes timestamps, and the last encode stop section at the end of each snapshot.

\vspace{0.3cm}

\textbf{Interval Graph encoding}
For this representation, we need to introduce a new property: the encoding length won't depend on the total number of events $e$, but on the total number of intervals $i$.  

As explained in section \ref{motivation}, an interval of edge existence corresponds to a period of time during which all possible observations if this edge are present. For instance, if observations occur every year, and the edge $e$ is observed in 2010,2011,2012 and 2013 but not in 2009 and 2014, then the four observations between 2010 and 2013 can be replaced by a single interval $[2010,2014]$.

As a consequence, we also define $t'\leq t$ the total number of different intervals endpoints, and $I^{t'}=\log_2(t')$.

\[
I^{IG}=mI^m+2iI^{t'}+mI^{t'}
\]
Where the first element is the cost of encoding the edges, the second element encodes the intervals, and the last encode stop sections as for link streams.

\section{Experiments}
\label{experiemnts}

To validate the relevance of our approach, we experiment with synthetic and real networks. In each experiment, we test with the original temporal resolution (on the left of figures), and then explore how aggregating at coarser temporal scales affects encoding costs. To create those aggregated version, we use non-overlapping sliding time-windows. To every unique time period, we associate an unweighted cumulative graph, such as an edge exists between two nodes of this graph if there is at least one interaction between those two nodes during the corresponding period.

All experiments can be checked and reproduced using an on-line notebook\footnote{\url{https://colab.research.google.com/github/Yquetzal/tnetwork/blob/master/article_encoding.ipynb}} 

\subsection{Synthetic networks}
\begin{figure}[h!]
  \centering
  \begin{subfigure}[b]{0.45\linewidth}
    \includegraphics[width=\linewidth]{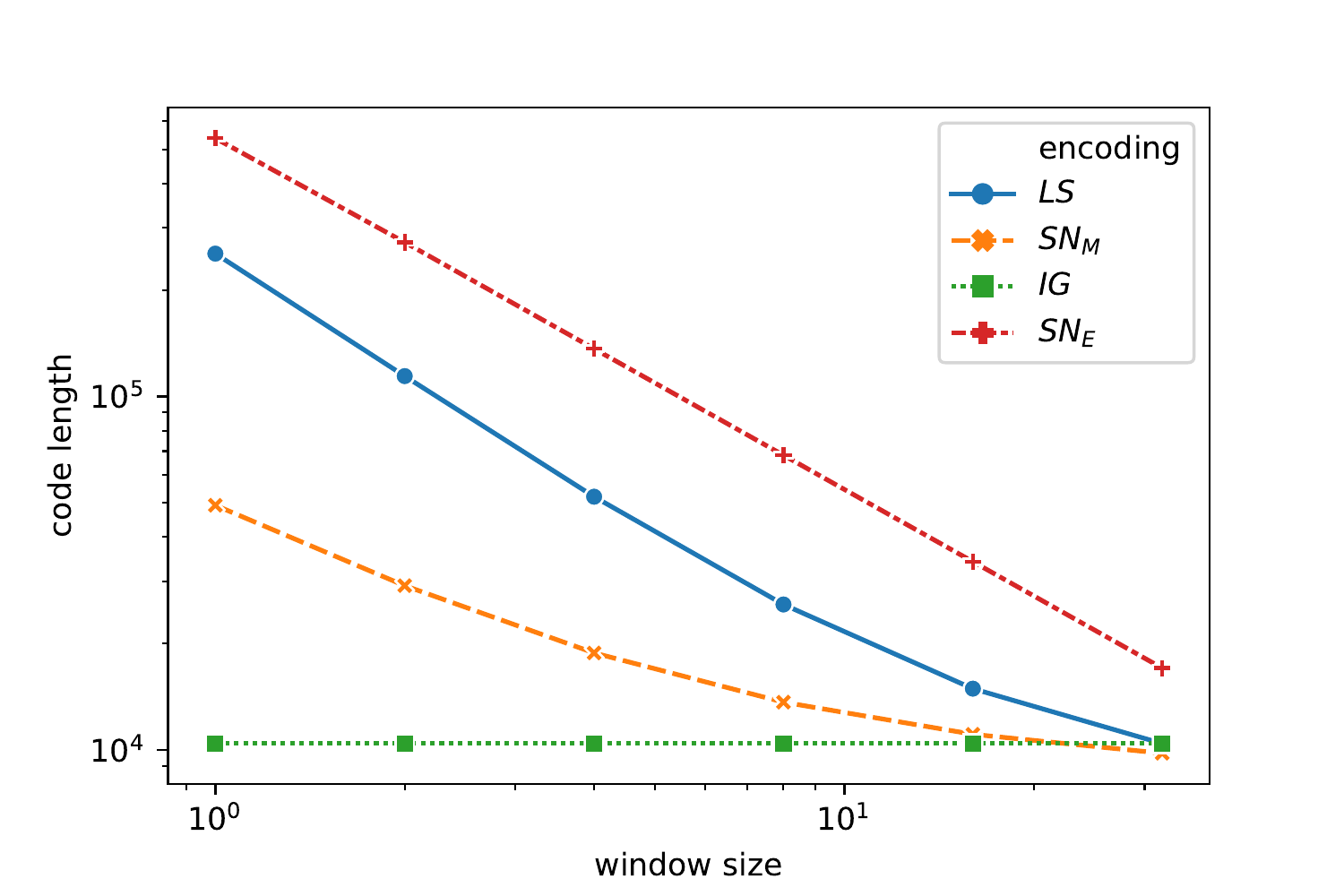}
    \caption{Stable Network}
  \end{subfigure}
  \begin{subfigure}[b]{0.45\linewidth}
    \includegraphics[width=\linewidth]{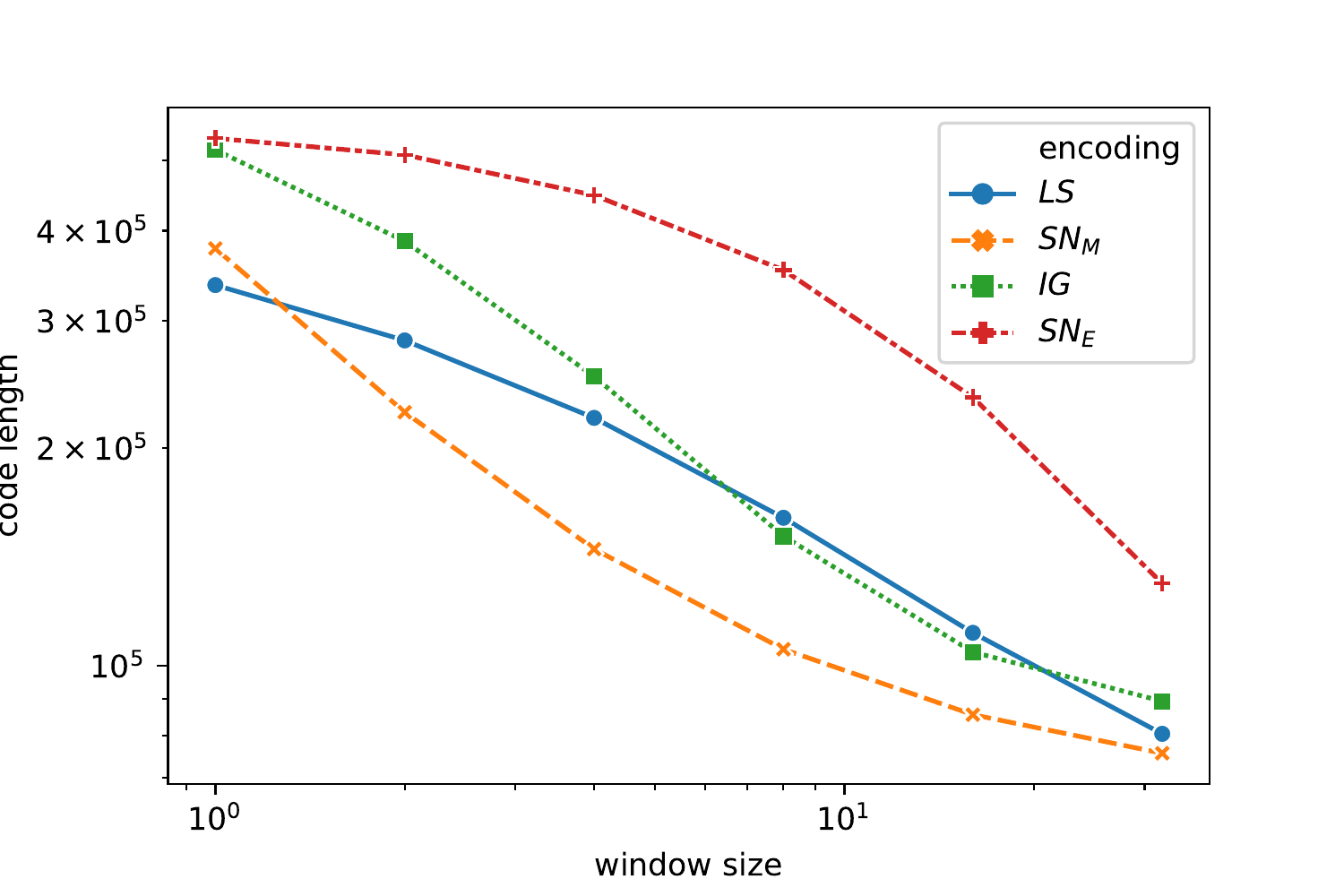}
    \caption{Independent snapshots, dense}
  \end{subfigure}
  
       \begin{subfigure}[b]{0.45\linewidth}
    \includegraphics[width=\linewidth]{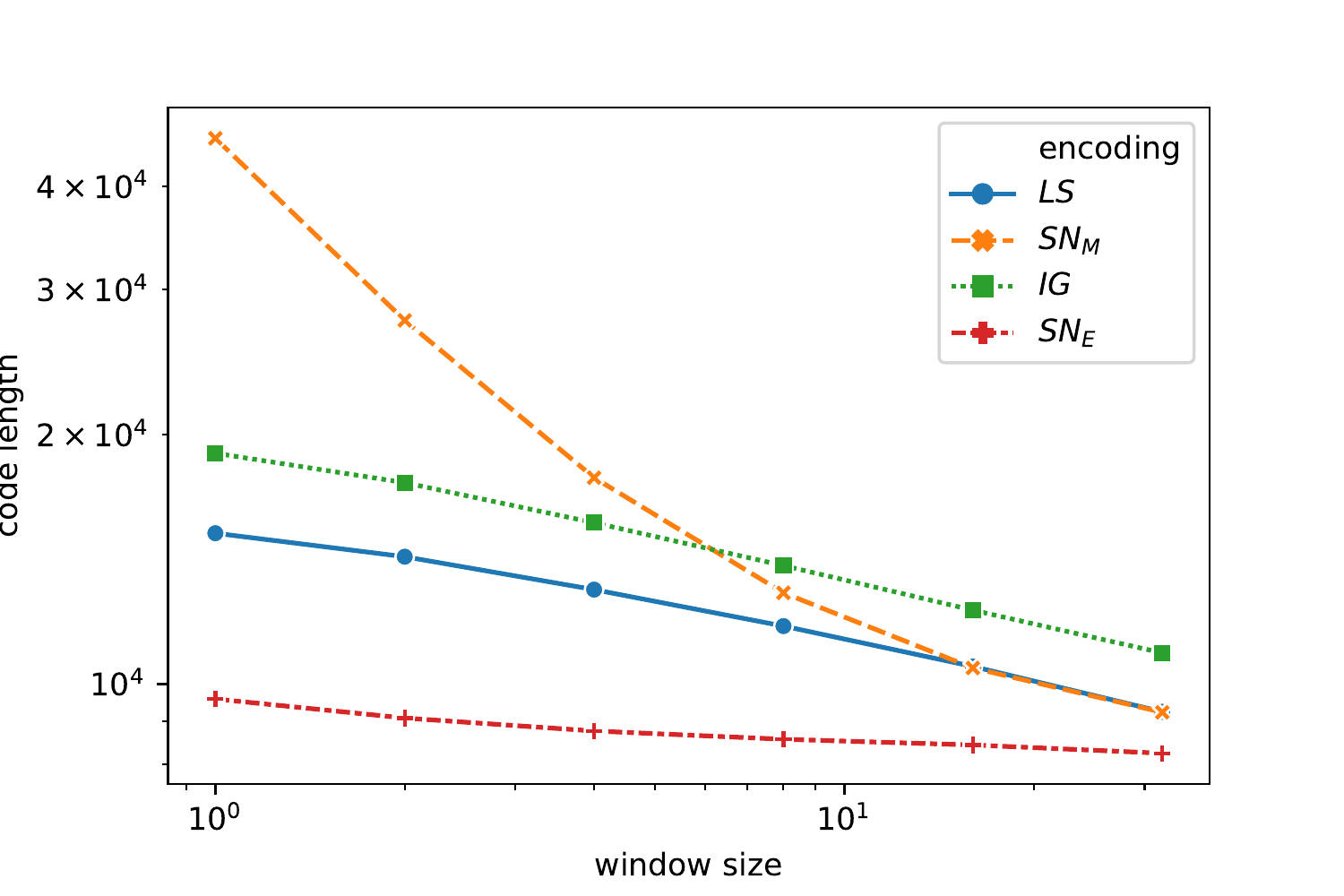}
    \caption{Independent snapshots, sparse}
  \end{subfigure}
     \begin{subfigure}[b]{0.45\linewidth}
    \includegraphics[width=\linewidth]{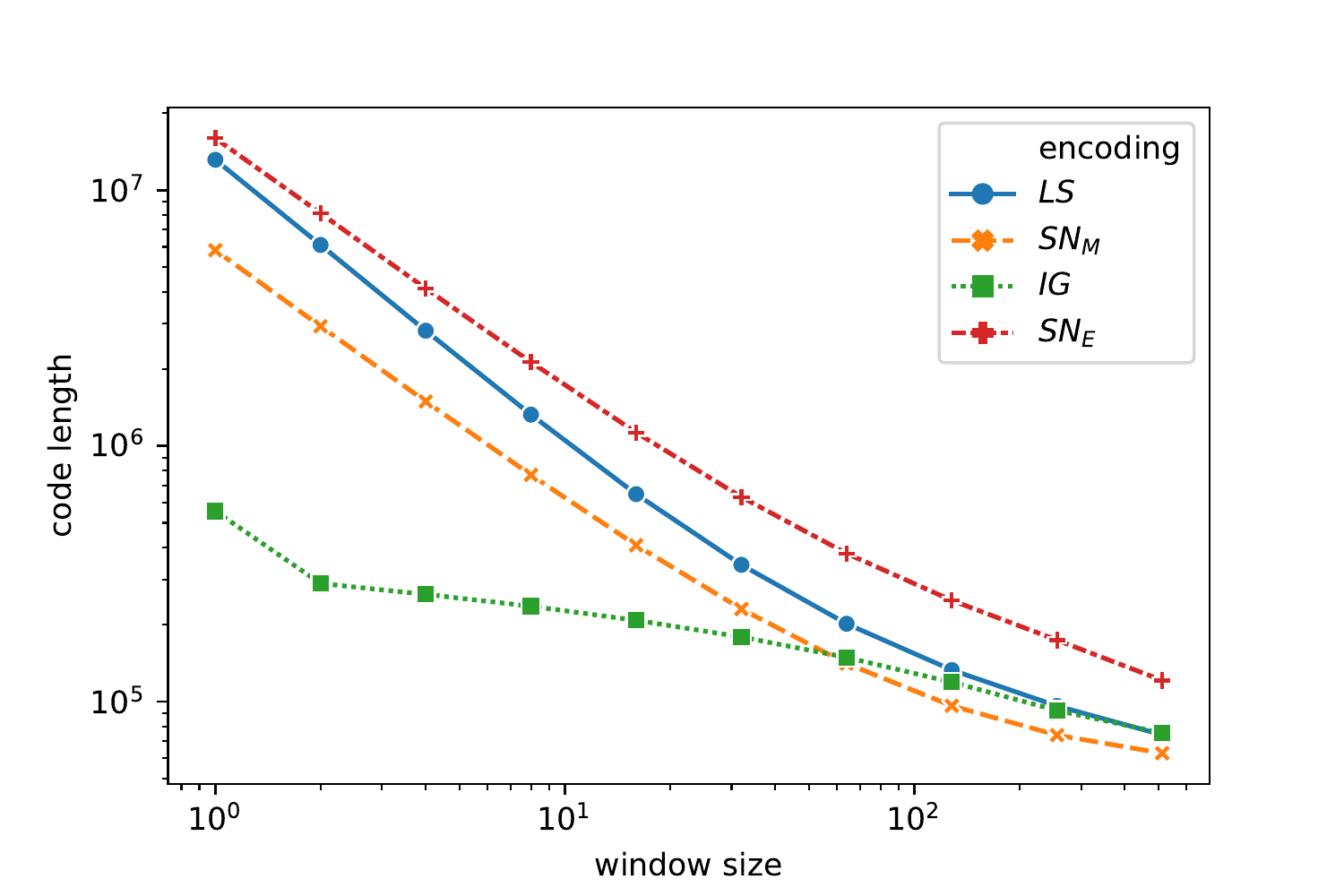}
    \caption{PEG Benchmark}
  \end{subfigure}
  
  \caption{Code length for synthetic graphs. We observe that the most efficient network representation depends strongly on the properties of networks.}
  \label{fig:synthetic}
\end{figure}

We generate three types of synthetic dynamic networks and compute encoding length on them for our four models. Results are synthesized in Fig. \ref{fig:synthetic}.

\vspace{0.3cm}

\textbf{Stable network.}
In the first experiment, we generate a single static Erd\H{o}s-Rényi random graph with 100 nodes and 640 edges and create a dynamic network composed of 64 identical snapshots. The properties of this temporal graph is therefore ($n=100,m=640,t=64,e=40960$). We can observe that the Interval Graph approach allows the most efficient encoding when there are at least two snapshots. This matches our expectation, since the network is perfectly \textit{stable} (long intervals). Link streams and $SN_E$ are the least efficient, because they repeat uselessly timestamps and edges, respectively.
\vspace{0.3cm}

\textbf{Independent snapshots, dense.}
In the second experiment, we generate a network with $t=64$ snapshots, each corresponding to a static Erd\H{o}s-Rényi random graph with the same property as previously (100 nodes, 640 edges). We observe in Fig. \ref{fig:synthetic} that the link stream representation is the most efficient without aggregation, but the matrix snapshot representation becomes more efficient as soon as we aggregate. This is due to the important density: with a total of $e=40960$ observed interaction for 4950 possible node pairs, most pairs are repeated multiple times.
\vspace{0.3cm}

\textbf{Independent snapshots, sparse.}
In this experiment, we generate a network with $t=64$ snapshots, each corresponding to a static Erd\H{o}s-Rényi random graph with 100 nodes and 10 edges. The number of observations ($e=640$) is now equivalent to the number of different edges $m$ observed in the first experiment. We observe that the $SN_E$ representation is now the most efficient, which is coherent with our expectation, since this representation is efficient when edge repetitions are rare.

\vspace{0.3cm}

\textbf{Progressively Evolving Graph Benchmark}
In the last experiment, we use an existing benchmark to generate a dynamic network. Published in \cite{cazabet2020evaluating}, this benchmark generates progressively evolving graphs with changing community structures. We use the standard generator used in \cite{cazabet2020evaluating}, yielding a temporal network with $n=92,m=4169,e=2007930,t=1961$. We observe that the interval graph representation is by far the most efficient on this network, which is due to the progressively evolving nature of the graphs: edges present in a period tend to be also present in the next. However, because there is a long term evolution and some random noise, the matrix representation $SN_M$ is not efficient.

\subsection{Experiments with real networks}

\begin{table}

\centering
\rowcolors{1}{White}{LightBlue}
\begin{tabular}{|l|l|l|l|l|l|l|l|}
\hline
Network & n   & m    & e      & t     & e/t & e/m & e/m/t(\%)\\ \hline
\hline
SP-HS \cite{10.1371/journal.pone.0107878}  & 180 & 2220 & 45047  & 11273 & 4 & 20.29 & 0.18 \\ \hline
SP-Hosp \cite{10.1371/journal.pone.0073970}& 75  & 1139 & 32424  & 9453  & 3.4 & 28.4 & 0.3 \\ \hline
SP-PS  \cite{10.1371/journal.pone.0023176} & 242 & 8317 & 125773 & 3100  & 40.6 & 15.1 & 0.49 \\ \hline
ENRON \cite{networkrepo}  & 150 & 1526 & 24694  & 14832 & 1.7 & 16.2 & 0.11 \\ \hline
Primates\cite{networkrepo} & 25  & 280  & 1340   & 19   & 70.5 & 4.8 & 0.25 \\ \hline
GOT   \cite{bost2016narrative}   & 338 & 939 & 20011 & 1031 & 19.4 & 21.3 & 2.07 \\ \hline
\end{tabular}
\caption{Summary of real networks properties. n: number of nodes. m: number of different edges. e: number of interactions. t: number of timesteps. e/t: average number of observations per timestemps. e/m: average number of observation per edge. e/m/t: average probability to observe an existing edge at a given step.}
\label{tab:networks}

\end{table}

\begin{figure}[h!]
  \centering
  \begin{subfigure}[b]{0.45\linewidth}
    \includegraphics[width=\linewidth]{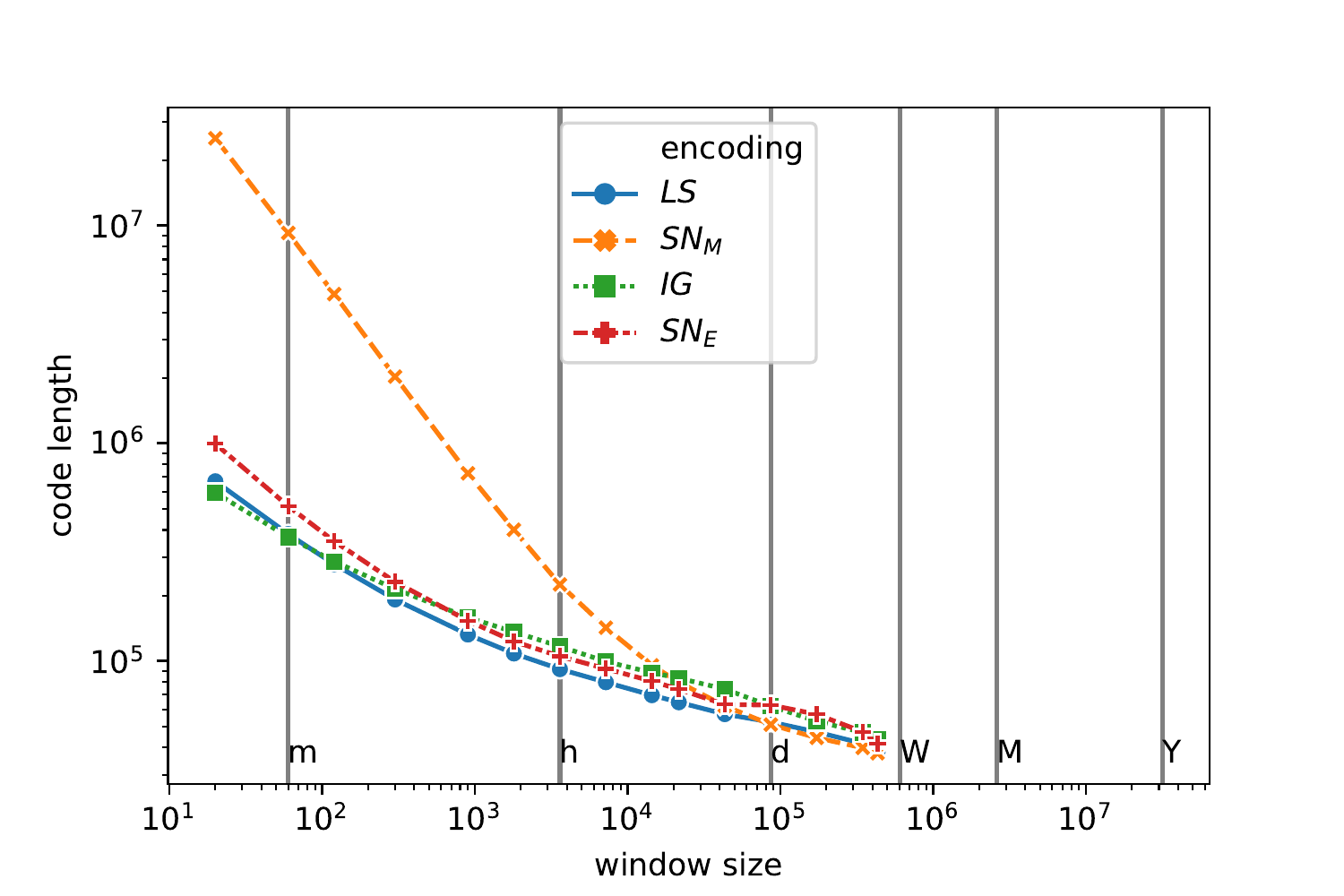}
    \caption{SP-HS}
  \end{subfigure}
  \begin{subfigure}[b]{0.45\linewidth}
    \includegraphics[width=\linewidth]{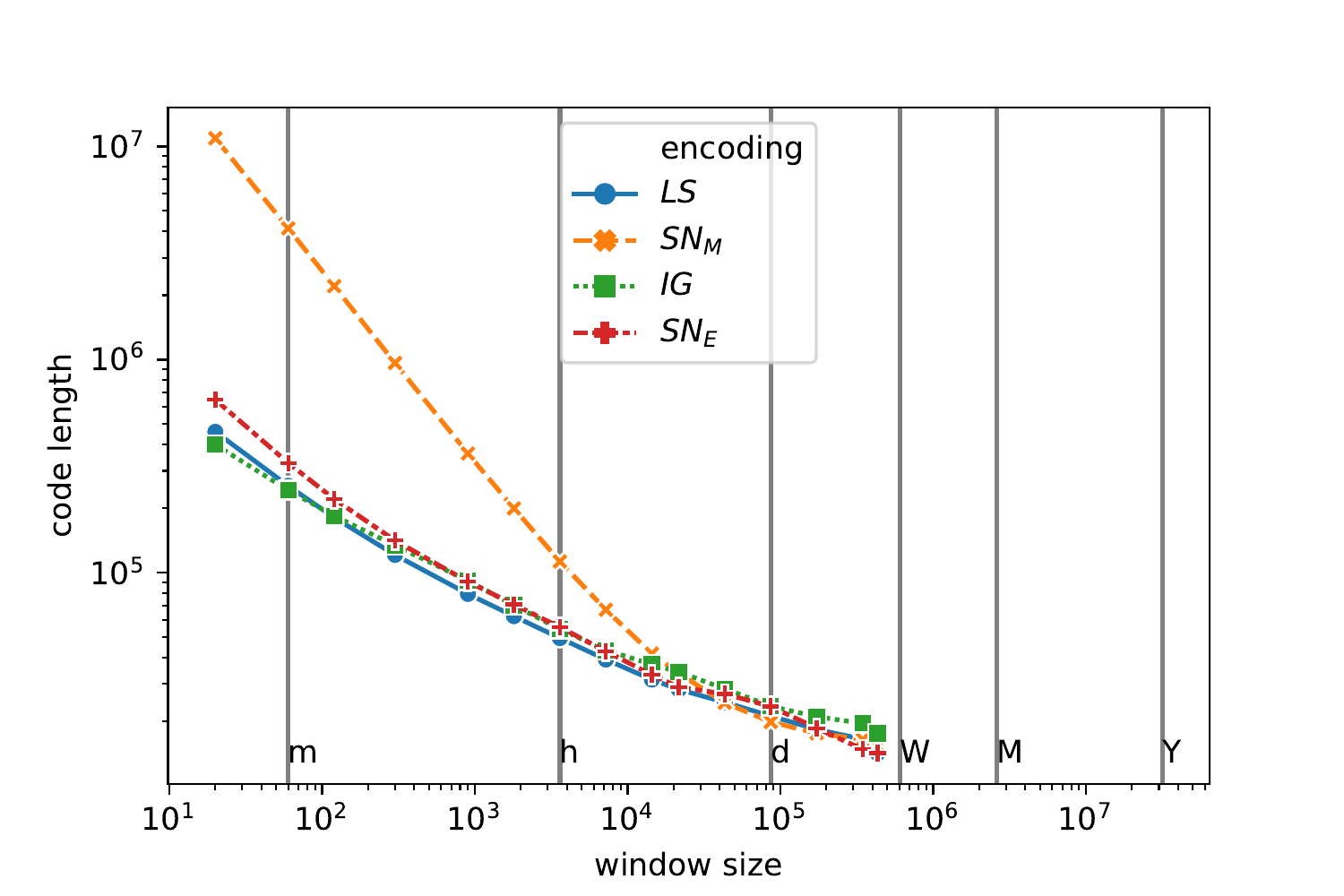}
    \caption{SP-Hosp}
  \end{subfigure}
  
   \begin{subfigure}[b]{0.45\linewidth}
    \includegraphics[width=\linewidth]{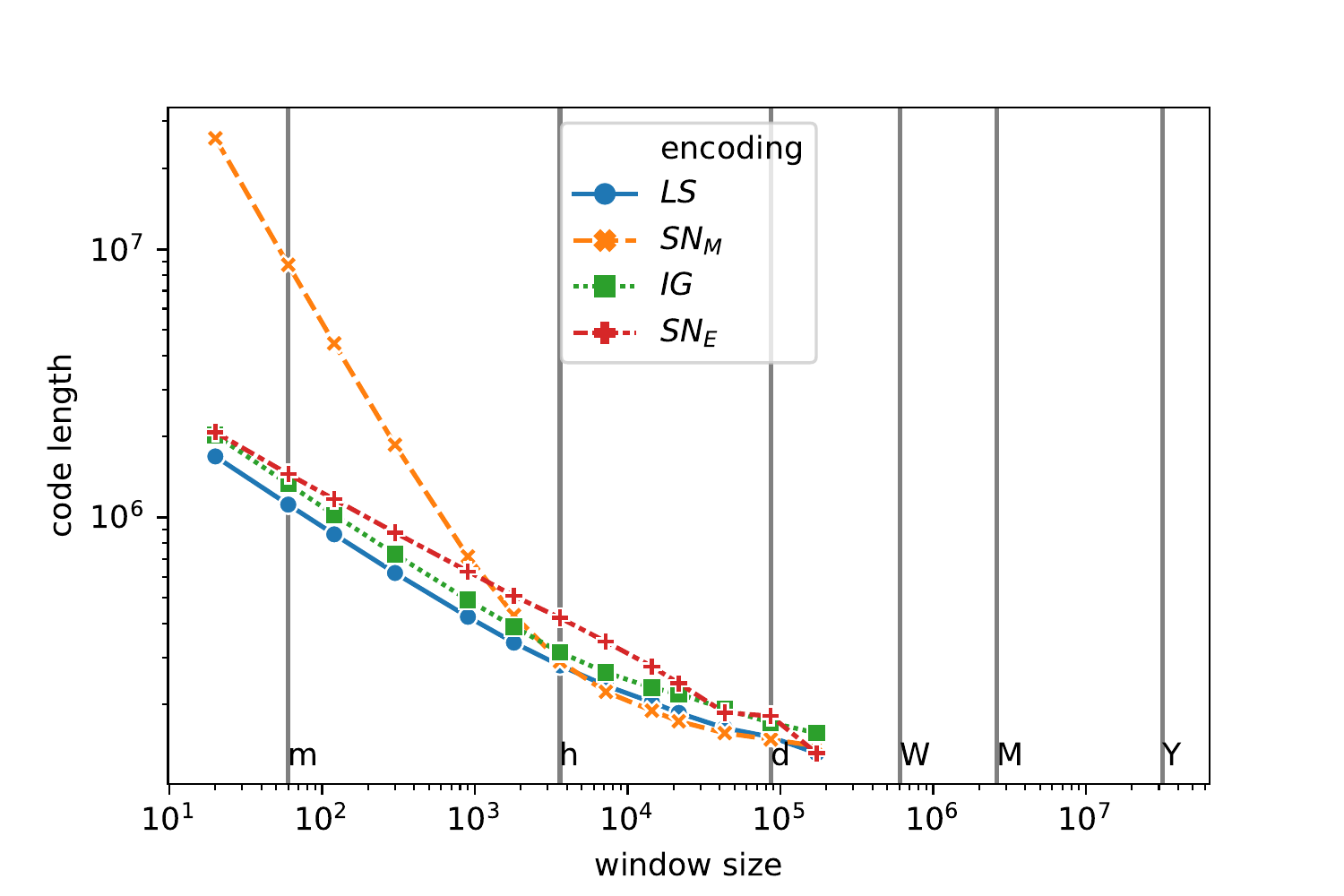}
    \caption{SP-PS}
  \end{subfigure}
  \begin{subfigure}[b]{0.45\linewidth}
    \includegraphics[width=\linewidth]{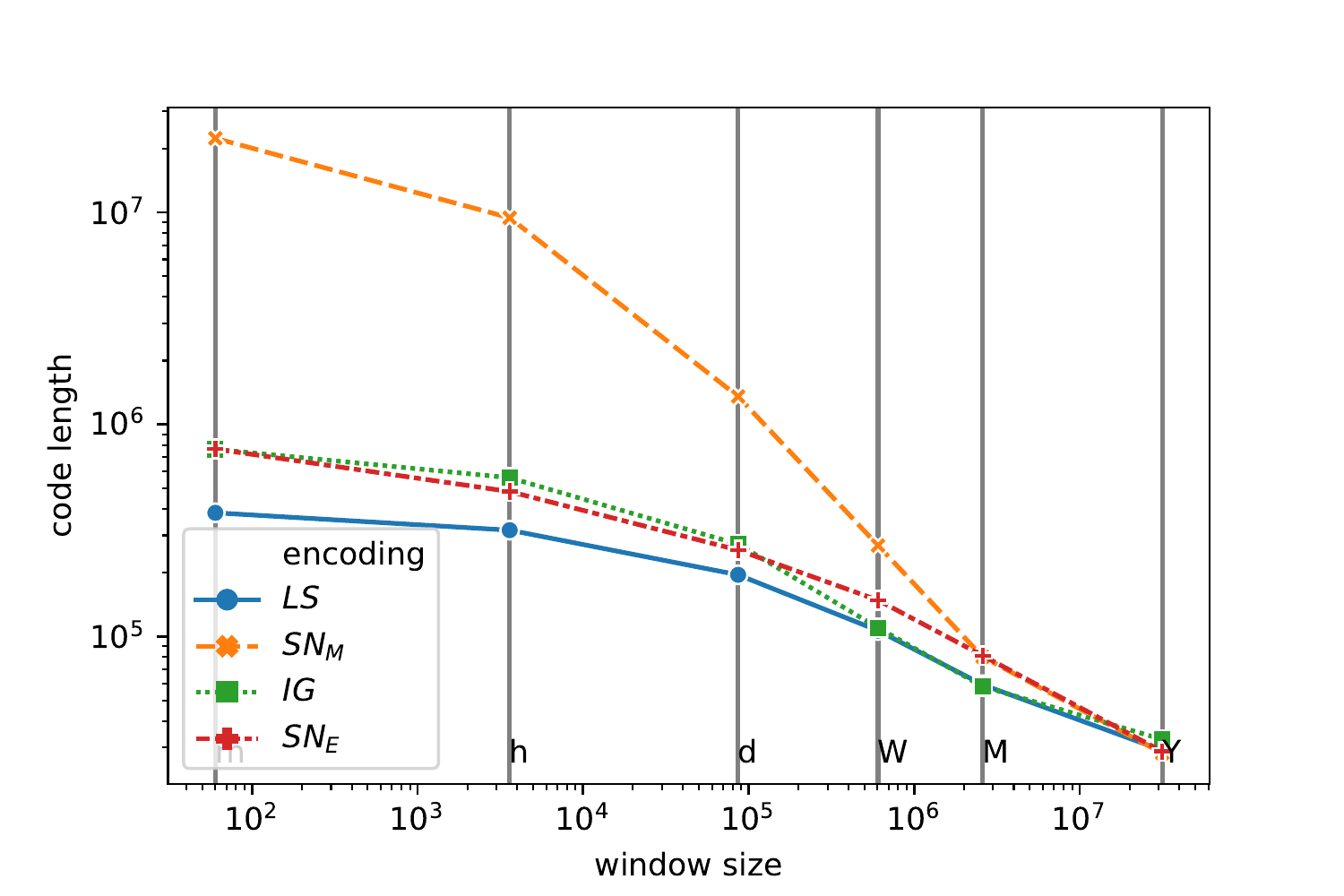}
    \caption{ENRON}
      \end{subfigure}

       \begin{subfigure}[b]{0.45\linewidth}
    \includegraphics[width=\linewidth]{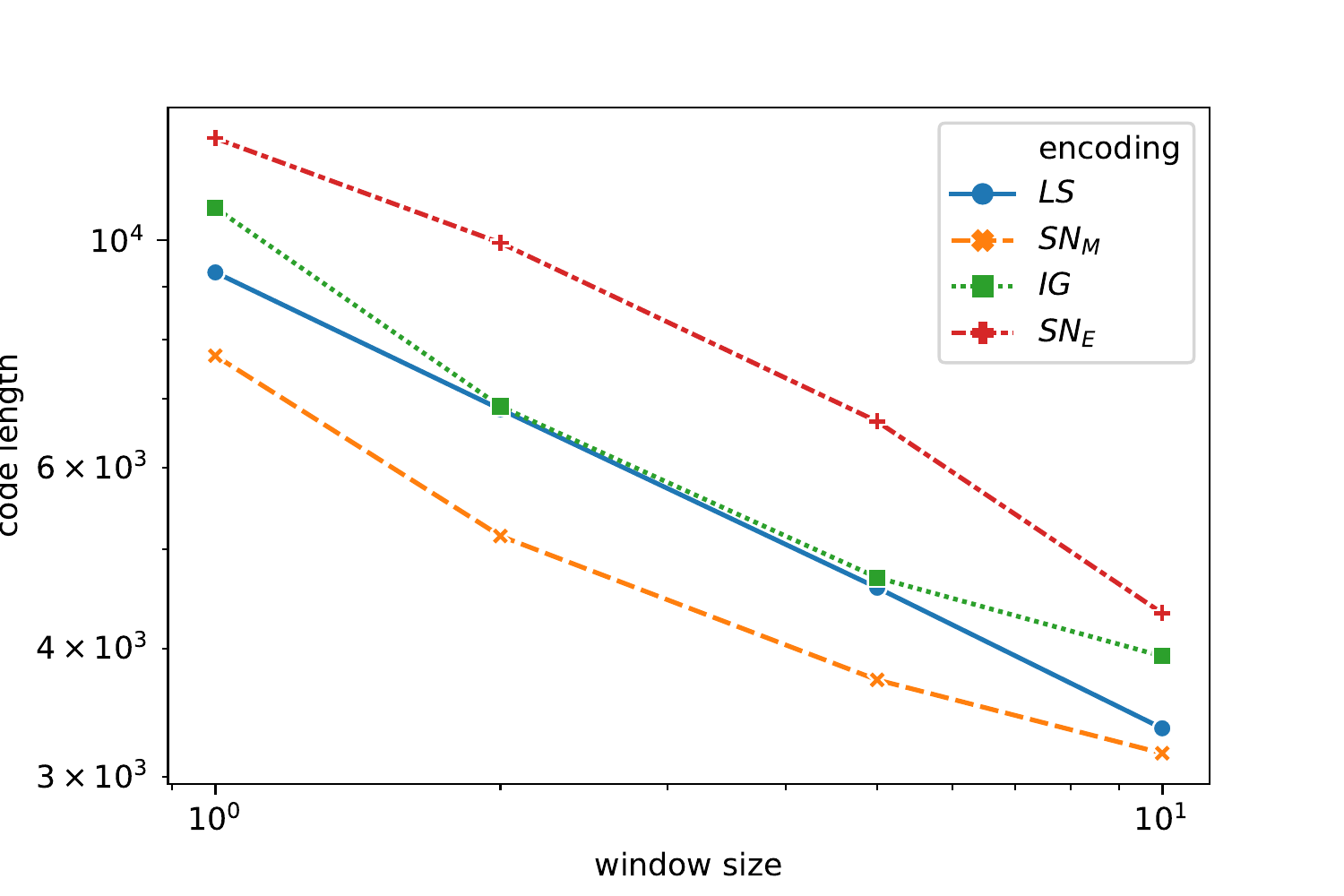}
    \caption{Primates}
  \end{subfigure}
  \begin{subfigure}[b]{0.45\linewidth}
    \includegraphics[width=\linewidth]{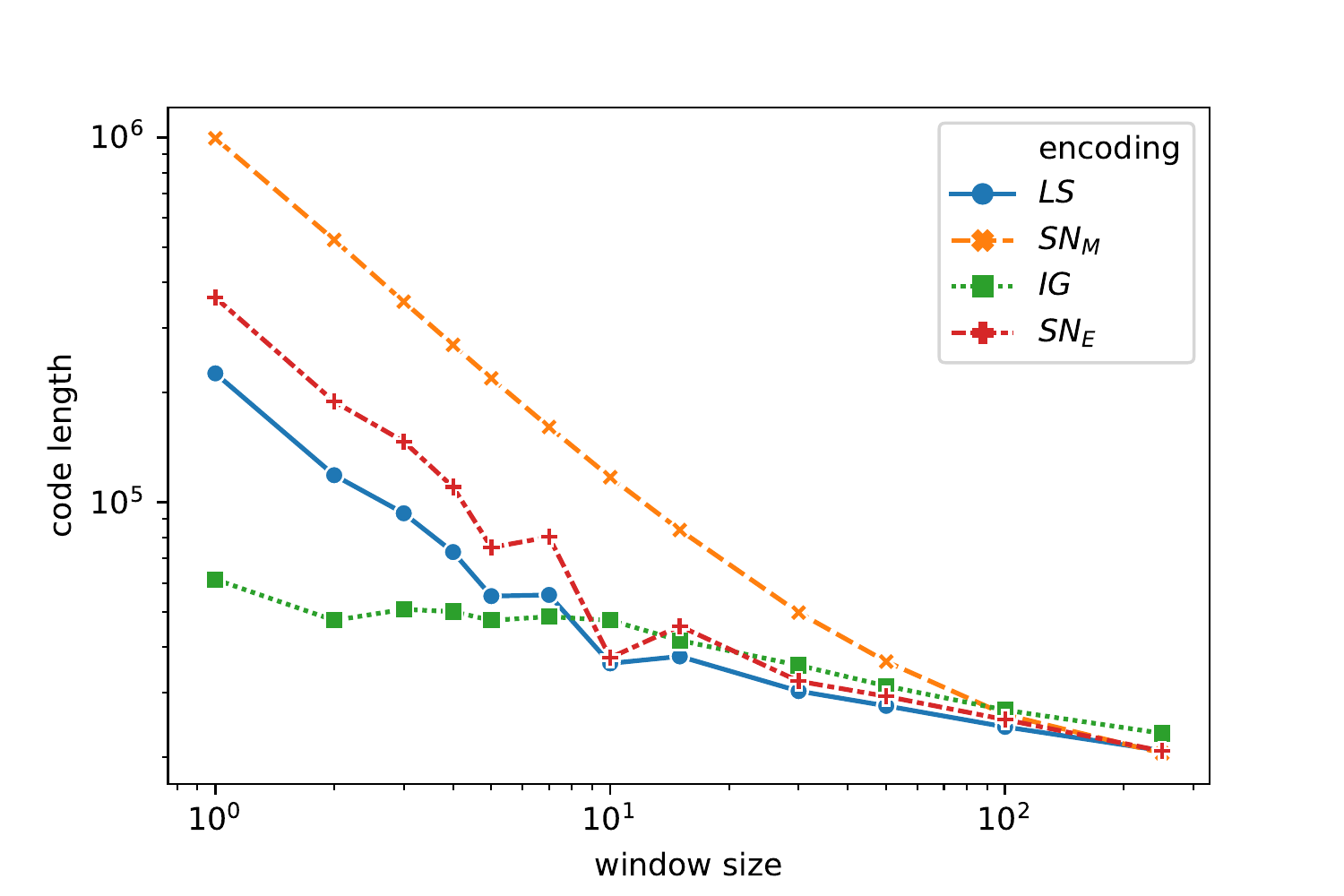}
    \caption{GOT}
  \end{subfigure}
  
  \caption{Code length for synthetic graphs. We observe that the most efficient network representation depends strongly on the properties of networks.}
  \label{fig:real}
\end{figure}

In this section, we apply the same procedure on temporal networks corresponding to real datasets. We summarize information on those networks in table \ref{tab:networks}. Three networks come from the SocioPatterns project \cite{barrat2013empirical}, SP-HS (High-School), SP-Hosp (Hospital) and SP-PS (Primary School). As already mentioned in section \ref{motivation}, they correspond to interactions collected between individuals every 20s. ENRON is a dataset of emails sent between employees. Temporal information is available at the level of the minute, over a period of about 3 years. Primates is a dataset of social interactions between primates, collected over 19 periods. GOT is a network of interactions between characters of a TV series (Game of Thrones) over several seasons, originally aggregated every 10 scenes.  

All datasets are public, and available either through their original paper or though the network data repository \cite{networkrepo}, as reported in Table \ref{tab:networks}.

Results of experiments are reported in Fig. \ref{fig:real}. When relevant, we indicate with vertical lines typical temporal scales (respectively, minute, hour, day, week, month, year).

The three sociopattern datasets seem to have similar profiles overall. For High School and Hospital datasets, Interval Graphs is the most efficient representation for the original timescale, while Link Streams become more efficient if we aggregate every 5 minutes, approximately. From this observation, we can infer that interactions tend to be maintained typically for no more than a few minutes. Each interval of observation thus becomes a single observation when aggregating, more efficiently represented as a Link Stream. For the Primary School, the Link Stream representation is the most efficient even for the original data, either because the data is more noisy or because there are more singleton observations.

Link Stream is also the most efficient representation for ENRON dataset, as expected due to the nature of the dataset: each email is stamped with the exact minute it was sent, and it is rather unlikely that emails sent on a particular minute form a well-defined graph, or that several emails are sent between the same individuals in successive minutes. Only when aggregating at a scale of weeks or months is the Interval Graph representation the most appropriate, and when aggregating every year a snapshot representation becomes relevant.

In the primate dataset on the contrary, the snapshot representation is the most appropriate: each timestamp correspond to a well formed graph, and relations are usually not stables from one snapshot to the next.

Finally, for the Game of Thrones dataset, Interval graphs seems to be clearly the most appropriate for the original data. However, by having a second look at the data, we realized that this is due to the way the dataset is provided. A smoothing window is used, and for each sequence of 10 scenes, the same average network is provided 10 times, i.e., snapshots 1-10 corresponds to the first 10 scenes and are identical. Using our method, we observe that if we aggregate every 10 scenes, thus removing this bias, the link stream approach becomes the most efficient.

\section{Conclusion}
\label{conclusion}

In this article, we have introduced a method to choose an appropriate representation for a temporal network, and validated its relevance on synthetic and real networks.

This method is implemented in the tnetwork python library to automatically select the right representation when loading a file, and to store more efficiently temporal networks. 

Beyond these practical aspects, choosing the most appropriate representation is essential to know how to handle a network and which algorithms or methods can be applied to it. In future works, we wish to analyze further how to select an appropriate aggregation scale to transform efficiently interaction datasets -- which seem to be the most frequent in real data -- into stable networks that can be analyzed as interval graphs or snapshots, while loosing as little temporal information as possible.

%
%
%
\bibliographystyle{plain}
\bibliography{refs.bib}

\begin{thebibliography}{10}

\bibitem{barrat2013empirical}
Alain Barrat, Ciro Cattuto, Vittoria Colizza, Francesco Gesualdo, Lorenzo
  Isella, Elisabetta Pandolfi, J-F Pinton, Lucilla Rav{\`a}, Caterina Rizzo,
  Mariateresa Romano, et~al.
\newblock Empirical temporal networks of face-to-face human interactions.
\newblock {\em The European Physical Journal Special Topics},
  222(6):1295--1309, 2013.

\bibitem{bost2016narrative}
Xavier Bost, Vincent Labatut, Serigne Gueye, and Georges Linar{\`e}s.
\newblock Narrative smoothing: dynamic conversational network for the analysis
  of tv series plots.
\newblock In {\em 2016 IEEE/ACM International Conference on Advances in Social
  Networks Analysis and Mining (ASONAM)}, pages 1111--1118. IEEE, 2016.

\bibitem{cazabet2010detection}
Remy Cazabet, Frederic Amblard, and Chihab Hanachi.
\newblock Detection of overlapping communities in dynamical social networks.
\newblock In {\em 2010 IEEE second international conference on social
  computing}, pages 309--314. IEEE, 2010.

\bibitem{cazabet2020evaluating}
Remy Cazabet, Souaad Boudebza, and Giulio Rossetti.
\newblock Evaluating community detection algorithms for progressively evolving
  graphs.
\newblock {\em Journal of Complex Networks}, 2020.

\bibitem{cazabet2019challenges}
Remy Cazabet and Giulio Rossetti.
\newblock Challenges in community discovery on temporal networks.
\newblock In {\em Temporal Network Theory}, pages 181--197. Springer, 2019.

\bibitem{coscia2012demon}
Michele Coscia, Giulio Rossetti, Fosca Giannotti, and Dino Pedreschi.
\newblock Demon: a local-first discovery method for overlapping communities.
\newblock In {\em Proceedings of the 18th ACM SIGKDD international conference
  on Knowledge discovery and data mining}, pages 615--623, 2012.

\bibitem{10.1371/journal.pone.0107878}
Julie Fournet and Alain Barrat.
\newblock Contact patterns among high school students.
\newblock {\em PLoS ONE}, 9(9):e107878, 09 2014.

\bibitem{gauvin2018randomized}
Laetitia Gauvin, Mathieu Genois, Marton Karsai, Mikko Kivel{\"a}, Taro
  Takaguchi, Eugenio Valdano, and Christian~L Vestergaard.
\newblock Randomized reference models for temporal networks.
\newblock {\em arXiv preprint arXiv:1806.04032}, 2018.

\bibitem{grunwald2007minimum}
Peter~D Grunwald and Abhijit Grunwald.
\newblock {\em The minimum description length principle}.
\newblock MIT press, 2007.

\bibitem{holme2012temporal}
Petter Holme and Jari Saram{\"a}ki.
\newblock Temporal networks.
\newblock {\em Physics reports}, 519(3):97--125, 2012.

\bibitem{latapy2018stream}
Matthieu Latapy, Tiphaine Viard, and Cl{\'e}mence Magnien.
\newblock Stream graphs and link streams for the modeling of interactions over
  time.
\newblock {\em Social Network Analysis and Mining}, 8(1):61, 2018.

\bibitem{matias2018semiparametric}
Catherine Matias, Tabea Rebafka, and Fanny Villers.
\newblock A semiparametric extension of the stochastic block model for
  longitudinal networks.
\newblock {\em Biometrika}, 105(3):665--680, 2018.

\bibitem{mucha2010community}
Peter~J Mucha, Thomas Richardson, Kevin Macon, Mason~A Porter, and Jukka-Pekka
  Onnela.
\newblock Community structure in time-dependent, multiscale, and multiplex
  networks.
\newblock {\em science}, 328(5980):876--878, 2010.

\bibitem{networkrepo}
Ryan~A. Rossi and Nesreen~K. Ahmed.
\newblock The network data repository with interactive graph analytics and
  visualization.
\newblock In {\em AAAI}, 2015.

\bibitem{10.1371/journal.pone.0023176}
Juliette Stehle, Nicolas Voirin, Alain Barrat, Ciro Cattuto, Lorenzo Isella,
  {Jean-Francois} Pinton, Marco Quaggiotto, Wouter {Van den Broeck}, Corinne
  Régis, Bruno Lina, and Philippe Vanhems.
\newblock High-resolution measurements of face-to-face contact patterns in a
  primary school.
\newblock {\em PLOS ONE}, 6(8):e23176, 08 2011.

\bibitem{10.1371/journal.pone.0073970}
Philippe Vanhems, Alain Barrat, Ciro Cattuto, Jean-Francois Pinton, Nagham
  Khanafer, Corinne Regis, Byeul-a Kim, Brigitte Comte, and Nicolas Voirin.
\newblock Estimating potential infection transmission routes in hospital wards
  using wearable proximity sensors.
\newblock {\em PLoS ONE}, 8(9):e73970, 09 2013.

\bibitem{viard2016computing}
Tiphaine Viard, Matthieu Latapy, and Cl{\'e}mence Magnien.
\newblock Computing maximal cliques in link streams.
\newblock {\em Theoretical Computer Science}, 609:245--252, 2016.

\end{thebibliography}

\end{document}